# Exploring Daily Emotional Well-Being through Personalized Data Analysis

Adriana Rodríguez Rodríguez
*Faculty of Engineering, University of Deusto, San Sebastián, Spain*

**Abstract**

This project analysed daily emotional well-being from a personalized perspective using records of emotional states and contextual information derived from daily life. It was based on the idea that well-being is not a fixed value, but varies from day to day, and can be better understood in relation to individual context and behaviour. The work is framed within the field of Data Science and focuses on emotional states at an individual level. The project focused on constructing an integrated daily dataset from multiple personal data sources and analysing emotional states. The distributions of emotional variables, their temporal evolution and the differences between users were studied, together with their variability over time. Personality and well-being questionnaires were also incorporated as stable references to better contextualize daily emotional results. A simple personalized recommendation approach was then developed by comparing similar days within each user, exploring how past contextual conditions relate to differences in the user's emotional state. Finally, machine learning techniques were used to evaluate whether daily contextual information contains signals that help distinguish between different types of days relative to a person's usual well-being. The project adopted an applied approach, with an exploratory scope focused on understanding patterns rather than building a fully predictive system. The results showed that emotional well-being is very personal and changes from day to day. General models had limited value, while personalized approaches captured these differences much better, highlighting the importance of analysing each person on their own terms.

## 1. Introduction

In recent years, the study of human behaviour and mental health has gained increasing relevance, partly due to the availability of digital data collected through personal devices such as smartphones [1]. Research on digital phenotyping (i.e., behavioural data collected via personal digital devices) and well-being increasingly combines information from active inputs such as questionnaires and self-reports with passive sensing signals such as mobile phone usage, physical activity, mobility, and sleep-related indicators [2].

These data are typically collected in a longitudinal manner, which makes it possible to observe how emotional and behavioural states evolve over time [3]. Questionnaires and self-reports provide subjective information about psychological state, while sensing measures offer more objective indicators of daily behaviour. Recent literature increasingly adopts multimodal approaches that combine self-report measures and passive sensing data to study behaviour and well-being [2].

Analysing emotional well-being using digital data is not straightforward, even when rich datasets are available. The information usually comes from different sources such as questionnaires, daily self-reports, and sensor-based measurements, and these sources do not share the same format or structure. In addition, real-world datasets often contain missing or incomplete values. Not all variables are recorded with the same regularity: some are measured many times per day, while others are collected only occasionally.

This work uses the StudentLife Dataset [4], a longitudinal dataset collected in a university setting that combines psychological questionnaires, daily self-reports, and multiple mobile sensing signals. Several relevant studies based on the StudentLife dataset have focused on different aspects, including dataset presentation, analysing behavioural correlations, and developing predictive models for specific mental-health-related outcomes [4] [5] [6]. The original StudentLife paper introduced the StudentLife sensing system and dataset and explored correlations between automatically sensed smartphone behavioural data and students' academic performance and mental well-being measures over an academic term [4]. Later works shifted the focus toward more specific predictive tasks, such as stress prediction using smartphone sensing data and personalized stress prediction using deep learning models [5] [6].

In contrast to previous works that mainly focus on specific prediction tasks, the present project explores a broader analytical perspective on daily emotional well-being using the StudentLife dataset. Instead of concentrating exclusively on predictive modelling, the work adopts a broader perspective that combines behavioural analysis and interpretable modelling approaches to better understand daily emotional well-being from behavioural and self-reported data.

The main objective of this project is to analyse daily emotional well-being using longitudinal multimodal behavioural and self-report data through a structured and interpretable analytical approach. To achieve this goal, the specific objectives are:

- To prepare and integrate daily data from questionnaires, self-reports, and behavioural records into a consistent analytical dataset.
- To perform a structured exploratory analysis of daily emotional data using descriptive statistics and visualizations.
- To measure emotional variability and stability at the individual level.
- To derive interpretable well-being and personality-related profiles from survey data.
- To explore personalized context-based recommendation strategies for daily emotional well-being.
- To analyse whether contextual and behavioural factors are associated with daily emotional well-being.
- To reflect on the ethical implications of working with personal data.

## 2. Data Preparation

### *2.1 Dataset*

The StudentLife dataset was collected as part of a longitudinal study conducted at Dartmouth College, where university students were monitored throughout an academic term [4]. The goal of the study was to capture daily behavioural and psychological patterns in real-world conditions.

The study was conducted during the Spring term of 2013, and students from a computer science course voluntarily participated by installing a smartphone application that continuously collected behavioural sensing data. The study initially involved 60 students. Alongside this automatic data collection, participants also completed ecological momentary assessments (EMAs), which consist of real-time self-reports collected during their daily life, as well as several validated psychological surveys administered at the beginning and end of the study period. In the analyses reported in the original paper, 48 students completed the study and were included in the final analysis [4].

### *2.2 Time handling and daily aggregation*

In the raw StudentLife dataset, most behavioural records are timestamped using UNIX epoch time in UTC. To enable consistent daily analysis, all timestamps were converted into a standard datetime format and adjusted to the America/New York time zone, which corresponds to the geographical context of the study.

The unit of analysis was defined at the user–day level, using (user_id, date) as the primary key. Based on this definition, the data were aggregated by user and by day using count, sum, mean, and other descriptive operations depending on the variable type. As a result, each row in the final analytical dataset represents a daily summary of behavioural and self-report information for each participant.

### 2.3 Handling missing records

A distinction was made between true absence of activity and absence of measurement. For behavioural variables expressed as counts, durations, or event totals, missing values were replaced with zero when the absence clearly indicated that no activity occurred on that day. In contrast, when a variable was not measured or no response was available, particularly in questionnaire-based data, the value was kept as missing (NaN). This distinction preserves the semantic difference between "zero activity" and "no information available," which is important for correct interpretation in subsequent analyses.

### 2.4 Data sources and variables

Each data source was processed according to its nature: behavioural logs, academic information, EMA responses, passive smartphone sensing data, and survey-based questionnaires. Survey-based measures were incorporated as individual-level characteristics describing relatively stable psychological and personal attributes of participants. Only pre-survey ("pre") responses were retained when both pre- and post-assessments were available. Sensing modules such as GPS, WiFi location, and conversation were excluded due to their higher processing complexity, limited interpretability, or redundancy with other contextual signals already considered. Table 1 summarizes the main variable blocks of the integrated dataset.

**Table 1.** Main Variable Blocks of the Integrated Dataset.

| Block | Source | Example variables |
|---|---|---|
| Behavioural logs | App usage, calendar, call log | total_apps_used, tiempo_uso_total (overall device activity), calendar_event_count, total_calls |
| Academic context | Deadlines, class EMA | num_deadlines_today, hours (study hours) |
| EMA emotional | Mood, stress | happy (1–4), stress_level |
| EMA contextual | Sleep, social | sleep_hours, sleep_quality, social_people |
| Passive sensing | Activity, Bluetooth, screen, lock, Wi-Fi, SMS | prop_active, prop_stationary, bt_unique_devices, dark_minutes, locked_minutes, unique_wifi, num_sms |
| Surveys (participant level) | BFI-44, Flourishing, UCLA Loneliness, PANAS, PSS-10, PHQ-9 | Big Five traits, flourishing, loneliness, positive/negative affect, perceived stress, depressive symptoms |

### 2.5 Final dataset and outcome variable

The final dataset consists of 4,753 daily user-level observations corresponding to 49 participants and 184 distinct days of monitoring. The final analytical dataset consisted of 77 columns, including 75 analytical variables and 2 identifier fields (user_id and date). Missing values are present across variables, but are more prevalent in EMA-related measures, since these data were only recorded when participants actively responded and are therefore not available for all user–day observations.

The target variable selected is the daily happiness score obtained from EMA mood reports. This variable was chosen because it directly reflects participants' self-reported emotional state in everyday situations and provides a concrete indicator of subjective well-being at the daily level. Happiness responses are recorded on an ordinal 4-point intensity scale ranging from 1 (low happiness) to 4 (high happiness).

After daily aggregation, a total of 220 user-day observations with valid happiness reports were available. These observations correspond to 38 participants and 47 distinct calendar days, confirming the irregular and sparse nature of EMA mood responses across users and time. The average happiness score across valid observations was 1.79 (SD = 0.83), with a median value of 2 on the 1–4 scale. The distribution is skewed toward lower values, with most reports concentrated at levels 1 and 2, while higher values (3 and 4) appear less frequently.

## 3. Methods

### *3.1 Baseline reference models*

Several simple baseline prediction strategies were evaluated using the mean absolute error (MAE), which measures the average absolute difference between the observed happiness values and the predicted values. First, a global mean baseline was defined by predicting the same value for all observations, equal to the overall mean happiness score. Second, to test whether using contextual and behavioural variables improves prediction compared to a simple baseline, a linear regression model was trained using sleep-related measures, a phone usage indicator reflecting overall daily device activity, perceived stress, and social interaction variables as predictors. Only days with complete data across the selected variables were retained, resulting in 65 observations from 13 participants. Third, to account for stable individual differences in reported happiness levels, a participant-specific baseline was computed by predicting each observation using that participant's own mean happiness across their available days.

### *3.2 Within-person day classification*

To account for individual differences in baseline happiness, each day was evaluated relative to the participant's own average happiness across available days. Days were then classified as better-than-usual or at-or-below-usual by comparing the reported happiness score (happy) with the individual's personal mean. Daily contextual variables were then aggregated by computing the average value of each variable within these two groups. The variables used were sleep duration, sleep quality, number of apps used, total smartphone usage time, social interaction, and perceived stress.

### *3.3 Emotional variability and stability*

For the variability analysis, users were required to have at least three valid daily EMA happiness reports. Two per-user summary metrics were used: mean daily happiness and the within-user standard deviation of happiness. The mean represents overall emotional level, while the standard deviation captures day-to-day variability.

### *3.4 Survey-based personality and well-being profiles*

The survey variables cover the Big Five personality dimensions (extraversion, agreeableness, conscientiousness, neuroticism, and openness) and psychological well-being measures such as flourishing, loneliness, affective balance from PANAS, perceived stress, and depressive symptoms measured through PHQ. Survey variables were aggregated at the user level before profile construction. Users with more than 20% missing responses across survey variables were excluded, and for the remaining data, missing values were imputed using the median of each variable. All survey variables were standardized using the z-score. A KMeans clustering algorithm was applied to the standardized survey variables, and different numbers of clusters were evaluated and compared using the silhouette coefficient.

### *3.5 Similarity-based contextual recommendation*

Each day is represented as a vector of 17 contextual and behavioural variables: sleep patterns (sleep_hours, sleep_quality), psychological state (stress_level), social interaction (social_people), smartphone activity (tiempo_uso_total, total_apps_used, locked_minutes, dark_minutes), mobility and activity patterns (prop_active, prop_stationary), digital environment indicators (unique_wifi, bt_unique_devices), communication behaviour (total_calls, num_sms), and academic workload (calendar_event_count, num_deadlines_today, hours). To ensure that similarity comparisons reflect deviations from an individual's usual behaviour, all features are standardized within each user using z-score normalization, as shown in Equation (1):

$$z = (x - \mu_{\text{user}}) / \sigma_{\text{user}} \qquad (1)$$

where x is the observed value for a given day, and μ_user and σ_user correspond to the mean and standard deviation of that variable for the same user. For a given user and a given day t, the recommendation system compares the current day only with previous days of the same user, avoiding the use of any future information. Similarity between days is computed using Euclidean distance, as shown in Equation (2):

$$d(x_t, x_i) = \sqrt{\Sigma_j (x_{i,j} - x_{t,j})^2} \qquad (2)$$

To ensure that similarity can be computed for all days, missing values were replaced with zero after feature standardization, which means that a missing value is treated as if the user had a typical value for that variable on that day. Only the K most similar days (nearest neighbours) are retained, with K set to 10. A day is labelled as better-than-usual if its reported happiness score exceeds the user's average happiness across the observation period, and among the K most similar past days, only those classified as better-than-usual are retained. For this subset, the system computes the average of the contextual variables and compares these values with those of the current day. To maintain interpretability, only the variables with the largest absolute differences are retained, producing a concise set of personalized recommendations.

### *3.6 Complementary rule-based analysis*

A complementary rule-based approach aims to identify patterns that appear multiple times in the user's data and are associated with better-than-usual emotional states. All selected features are normalized within each user using z-score standardization and discretized into three categories: values below −0.5 are labelled as LOW, values above 0.5 as HIGH, and values in between as NORMAL. The selected variables are stress_level, tiempo_uso_total, prop_active, prop_stationary, total_calls, num_sms, unique_wifi, calendar_event_count and num_deadlines_today. A minimum support of 4 means that a condition must appear in at least four days of the user's data, a minimum confidence of 0.60 means that at least 60% of the days where the condition appears must correspond to better days, and the maximum rule size is limited to two variables. The rules are extracted separately for each user and ordered according to their confidence and support.

### *3.7 Identifying factors associated with better-than-usual days*

A binary outcome variable called better_day is defined: a day is labelled as better_day = 1 if its happiness value is higher than the participant's average happiness, and better_day = 0 otherwise. The contextual variables were organized into several feature groups: academic context (calendar events, deadlines, and study hours), sleep-related factors (sleep duration and perceived sleep quality), psychological factor (perceived stress level), social interaction (number of people interacted with, nearby Bluetooth devices, calls, and SMS activity), phone behaviour (overall device activity, number of applications used, screen inactivity periods, and device locking time), and mobility and activity (activity proportions and number of detected Wi-Fi networks). In addition, a final configuration including all available contextual variables was also analysed. Contextual variables were normalized within each user using z-score standardization.

The split was performed at the user level using a group-based strategy (implemented using GroupShuffleSplit), so that all observations from a given participant were assigned either to the training set or to the testing set, but never to both simultaneously. The training set contained 166 daily observations from 30 users, while the test set contained 54 daily observations from 8 users. Missing values were handled within the modelling pipeline using median imputation calculated using only the training data. A Random Forest classifier was implemented using 300 decision trees, with a maximum depth of 8 and a minimum of 5 observations per leaf, and class balancing was applied. Model performance was primarily assessed using ROC-AUC, complemented by accuracy, the confusion matrix, and the classification report. Feature importance was computed using permutation importance, measured as the average decrease in ROC-AUC, with the permutation process repeated 20 times for each feature.

## 4. Results

### *4.1 Baseline models*

The global mean baseline resulted in a mean absolute error (MAE) of 0.71. The regression model resulted in a prediction error similar to the global mean baseline. This suggests that although daily behavioural and contextual signals are related to emotional variation, their predictive power remains limited in this dataset, possibly due to the small number of complete observations and the irregular nature of EMA responses. The participant-specific baseline reduced the prediction error to an MAE of 0.53, substantially improving upon both the global mean baseline and the contextual linear regression model. The comparison across baseline approaches indicates that between-person differences explain a large proportion of the variance in happiness scores.

### *4.2 Within-person analysis of daily happiness*

Across all observations, 72 days were classified as better-than-usual and 148 days as at-or-below-usual, indicating that days with higher-than-typical happiness occurred less frequently than neutral or lower days. On better-than-usual days, participants showed higher average sleep duration, greater mobile phone engagement both in overall device activity and number of apps used, as well as slightly higher perceived stress levels and social interaction. Sleep quality, however, was marginally lower on better-than-usual days (Table 2).

**Table 2.** Mean Contextual Values on Better-Than-Usual and At-or-Below-Usual Days.

| Variable | Better-than-usual | At-or-below-usual |
|---|---|---|
| Sleep duration (hours) | 8.30 | 7.81 |
| Overall device activity | 1313 | 919 |
| Number of apps used | 16.15 | 13.34 |
| Perceived stress level | 2.80 | 2.36 |
| Social interaction | 3.63 | 3.34 |
| Sleep quality | 1.65 | 1.92 |

Taken together, these patterns suggest that days with higher-than-typical happiness are associated with higher levels of measured daily activity and device usage indicators, rather than simply with lower demands or greater rest.

An example participant (User 59) was selected, as this participant had the highest number of days with complete observations across the selected variables. This user provided 43 days with valid happiness EMA reports. Based on the participant's personal mean happiness score (2.06), days were classified into 15 better-than-usual days and 28 at-or-below-usual days. The results show that better days are associated with higher stress levels (+1.09), slightly longer sleep duration (+0.17 hours), more apps used (+0.44), and a higher proportion of active time (+0.009). In contrast, sleep quality appears slightly lower on better days (−0.27). Overall, these differences suggest that improved emotional states are not necessarily linked to more relaxed or optimal conditions, but rather to specific combinations of contextual factors that vary within the individual.

### *4.3 Individual emotional variability and stability*

After applying the minimum observation criterion, 21 users were retained for the variability analysis. Figure 1 shows the relationship between mean happiness and emotional variability across users. The scatter plot highlights substantial differences in stability: users with similar mean happiness can exhibit very different levels of variability. Some users exhibit almost no day-to-day variation, while others show pronounced fluctuations. Higher average happiness does not necessarily correspond to greater stability. These results indicate that emotional level and emotional stability capture different aspects of well-being.

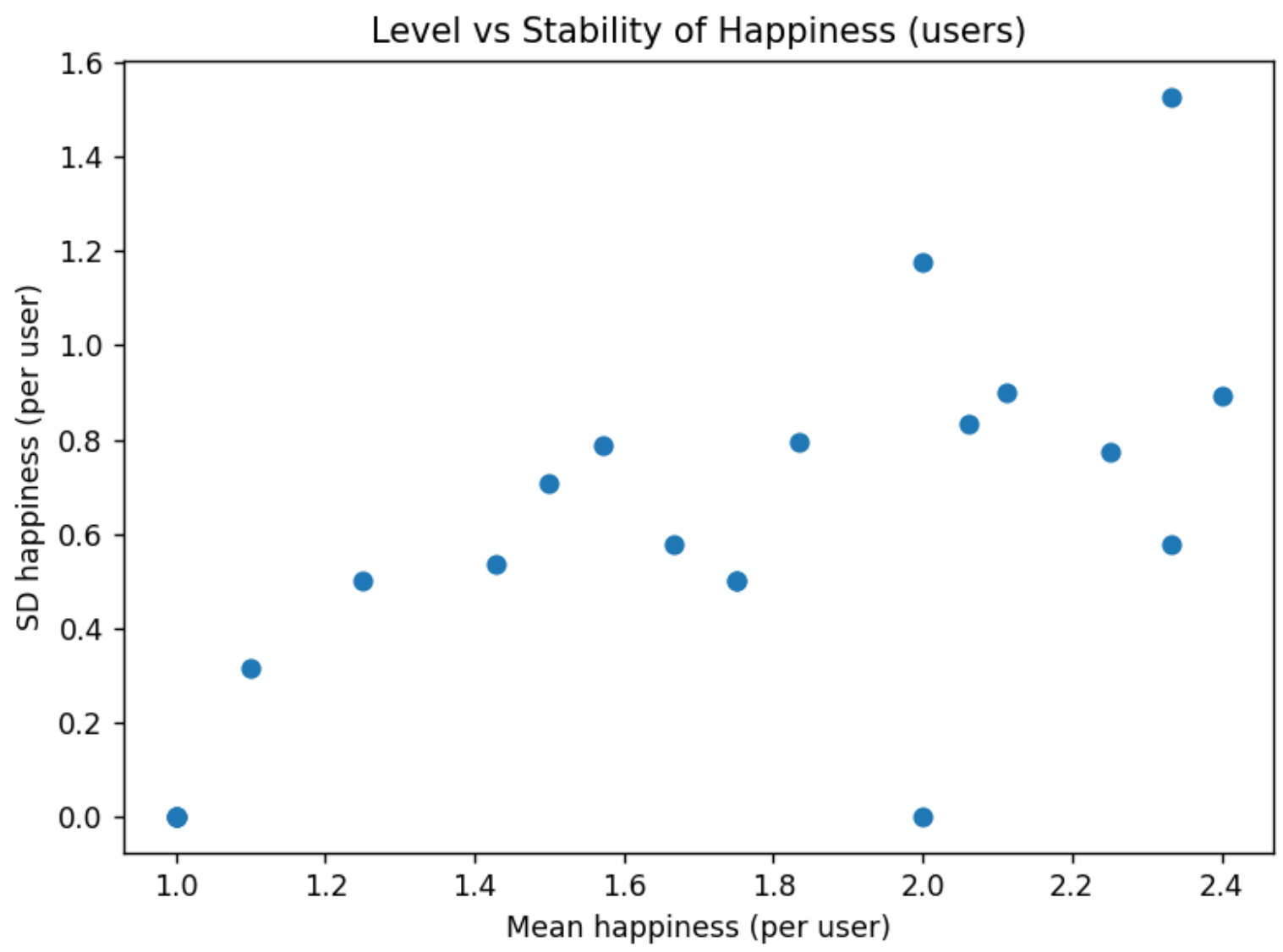

**Figure 1.** Mean Happiness vs Emotional Variability Across Users.

### *4.4 Survey-based personality and well-being profiles*

The configuration with two clusters produced the highest silhouette score (0.19), compared to 0.15 for k = 3 and 0.12 for k = 4, and was therefore selected as the final profile solution. Although the absolute value is relatively low, this is expected given the continuous and overlapping nature of psychological variables. The clustering procedure divided the 46 users remaining after the data filtering step into two groups: Profile 1 (n = 24) and Profile 2 (n = 22).

Profile 1 shows higher scores in extraversion, agreeableness, and conscientiousness, together with lower neuroticism. This group also presents higher flourishing and positive affect, and lower levels of loneliness, perceived stress, negative affect, and depressive symptoms. Overall, this profile reflects a psychologically resilient and higher well-being pattern. Profile 2 presents the opposite tendency, with lower scores in extraversion, agreeableness, and conscientiousness, and higher neuroticism. Users in this group report lower flourishing and positive affect, and higher loneliness, stress, negative affect, and depressive symptoms. This profile represents a more vulnerable well-being pattern (Figure 2). These profiles should be interpreted as dominant psychological tendencies rather than fixed personality types.

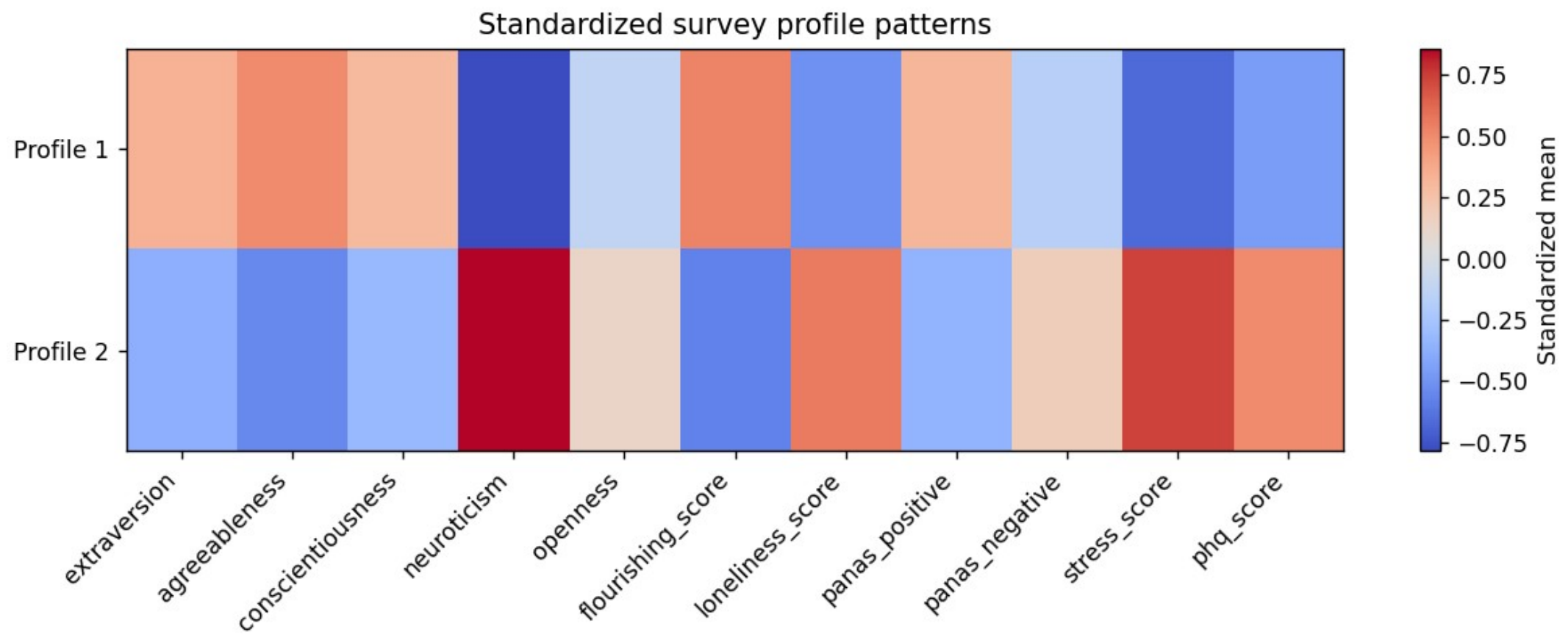

**Figure 2.** Heatmap of Standardized Survey Means by Profile.

After merging the profile labels back into the daily dataset, the analysis kept 4,443 daily observations from the same 46 users. Profile 1 shows slightly higher average levels of positive emotional states compared to Profile 2, and also reports, on average, a higher number of daily social interactions. In contrast, Profile 2 presents lower mean happiness and slightly higher perceived stress. Profile 2 exhibits higher variability in positive emotional variables such as happiness and general well-being. Both profiles report similar average sleep duration, while Profile 2 exhibits substantially higher overall device activity and a larger number of applications used. Taken together, these analyses indicate that survey-based personality profiles are associated with meaningful, although moderate, differences in daily emotions and behaviours.

### *4.5 Similarity-based contextual recommendation*

An illustrative example is presented for User 59 on 2013-06-07, a day with a reported happiness score of 2.0, indicating a relatively low emotional state for this individual. For this user, a total of 42 historical days were available for comparison prior to the target date. Among the 10 nearest neighbours, 4 days correspond to above-average emotional states for that user and are therefore labelled as better-than-usual. By comparing the values of the current day with the average values of similar better days, the comparison yields the following patterns: higher overall device activity (tiempo_uso_total: +1.05), longer sleep duration (sleep_hours: +0.88), more social interaction (social_people: +0.77), lower academic workload (hours: −0.73), more stationary time (prop_stationary: +0.60) and lower physical activity (prop_active: −0.52). Importantly, these patterns are user-specific and should not be interpreted as universal behavioural recommendations, but rather as personalized insights derived from the individual's own historical patterns.

In the complementary rule-based analysis, for user 59, a total of 43 daily observations with valid happy values are available, and the proportion of better days is 0.349. For this user, the results suggest that better days are often associated with specific combinations of daily conditions rather than a single factor. Patterns involving stress level, device activity, and daily workload appear frequently among the top rules. A second example is shown for user 49, who has 16 available days with happy values and a better-day rate of 0.438. In this case, the extracted rules suggest that better days are often associated with lower stress levels and moderate or balanced activity patterns for this user.

### *4.6 Factors associated with better-than-usual days*

Among the individual feature groups, academic context showed the strongest overall signal, with a ROC-AUC of 0.648. Sleep-related variables also showed relatively strong performance, with a ROC-AUC of 0.630. By contrast, the psychological feature group, represented only by stress level, showed very limited performance, with a ROC-AUC of 0.517, close to random discrimination. The social interaction group also produced low performance (ROC-AUC = 0.549). Phone behaviour produced intermediate results, with a ROC-AUC of 0.599. Mobility and activity variables showed relatively weak performance (ROC-AUC = 0.540). Finally, the model including all contextual variables achieved the highest accuracy (0.667) and a ROC-AUC of 0.612 (Table 3). However, the model still showed greater success in identifying non-better days than better-than-usual days.

**Table 3.** Model Performance (ROC-AUC) Across Feature Groups.

| **Feature group** | **ROC-AUC** |
|---|---|
| Academic context | 0.648 |
| Sleep-related factors | 0.630 |
| All features | 0.612 |
| Phone behaviour | 0.599 |
| Social interaction | 0.549 |
| Mobility and activity | 0.540 |
| Psychological factor | 0.517 |

Within the academic context model, calendar_event_count shows a substantially higher importance than the rest (0.118928). In the sleep-related model, sleep_quality emerges as the most influential variable (0.131605), with a notably higher importance than sleep_hours (0.054972). In the phone behaviour model, total_apps_used (0.057457) and dark_minutes (0.056818) were the most influential variables.

When considering the model trained with all features, the importance ranking changes notably: variables such as sleep_quality and stress_level appear as the most influential, while several variables exhibit near-zero or negative importance values (Figure 3). No single feature dominates across all models, highlighting the complexity of predicting better-than-usual emotional days.

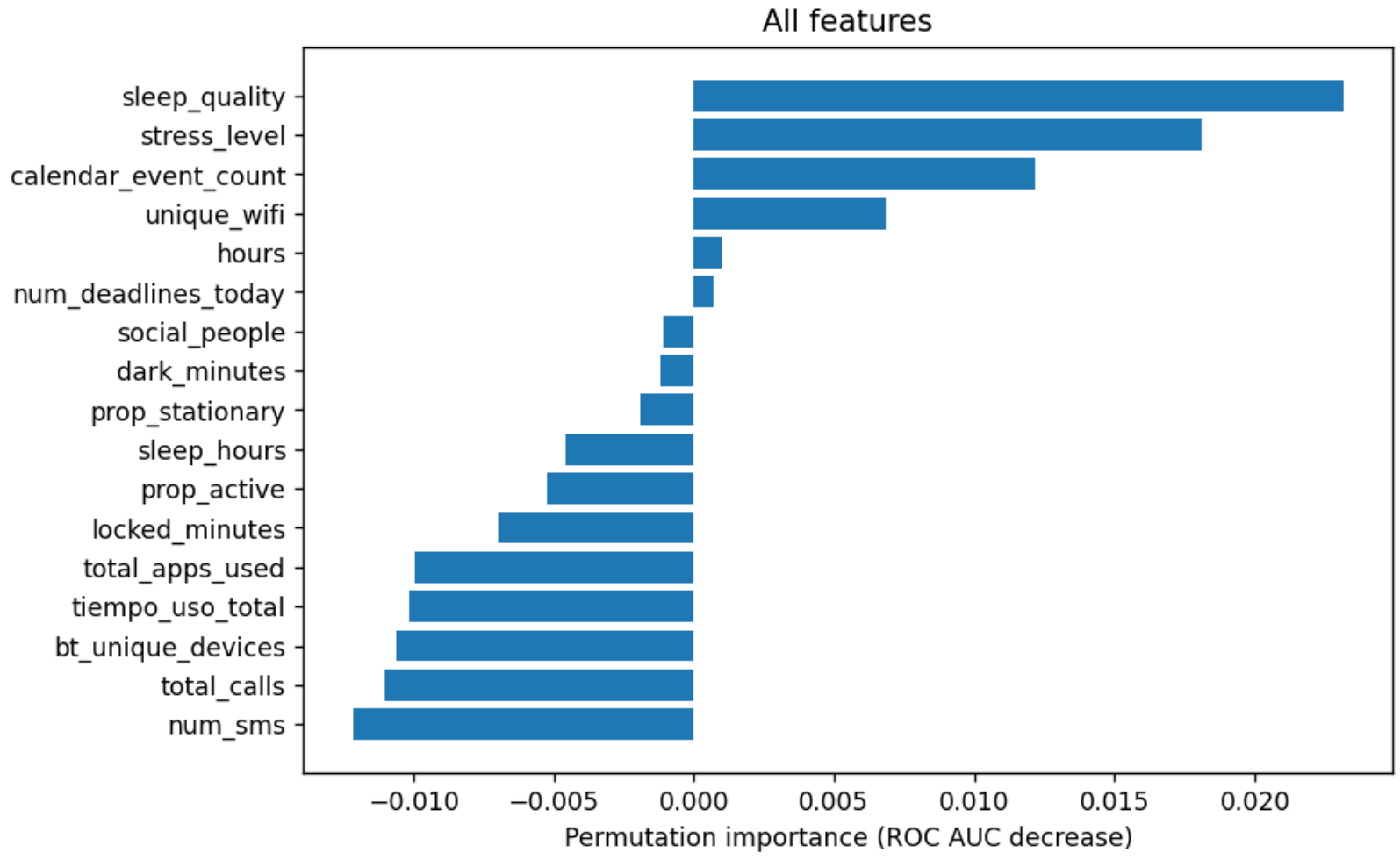

**Figure 3.** Feature Importance in the All-Features Model.

## 5. Discussion

Taken together, the results of this project suggest a clear idea: emotional well-being is highly individual, context-dependent, and difficult to fully capture using general models. The exploratory analysis showed that general predictive approaches provide limited value, while personalized baselines offer a more meaningful way to interpret emotional states. This is reinforced by the analysis of emotional variability, which shows that individuals with similar average happiness can still experience very different day-to-day emotional patterns.

The survey-based analysis adds another layer to this interpretation. By introducing stable psychological measures, it becomes possible to better capture differences between individuals. Building on these findings, the similarity-based approach provides a more intuitive and personalized way of analysing emotional well-being. Instead of trying to generalize across users, it focuses on each individual's own past experiences, allowing patterns to emerge in a more natural and interpretable way.

At the same time, the machine learning analysis highlights the limitations of relying solely on contextual variables to identify better-than-usual emotional days. The moderate performance of all models suggests that the selected variables do not fully capture the complexity of emotional states. The fact that models perform better at identifying typical days than positive deviations further reinforces the idea that emotionally better days are more subtle and harder to detect.

### *5.1 Limitations*

One of the main limitations of this study is related to the availability and quality of the happiness data. Daily emotional states are measured using EMA reports, which are collected irregularly and are not available for all users and days. As a result, the analysis is based on a relatively small subset of observations, which may limit the robustness of the findings. In addition, missing data are present across several variables, particularly in self-reported measures.

Survey-based measures represent a static snapshot of the individual and do not capture possible changes over time, and they rely on self-reported information, which may be affected by subjective bias. In the similarity-based approach, missing values are replaced with 0 after normalization, which may reduce the model's ability to distinguish between different situations. Finally, the dataset used for classification is relatively small and imbalanced, and the definition of better-than-usual days as a relative outcome introduces ambiguity in the classification task. As a result, the findings should be interpreted as exploratory and indicative rather than conclusive.

### *5.2 Future work*

Future work could focus on using larger and more complete datasets, with more consistent emotional reports across users and time. Future work could also explore alternative strategies for handling missing data, such as K-nearest neighbours (KNN) imputation. Future studies could incorporate multiple emotional dimensions, and sequence-based machine learning models, such as recurrent neural networks (e.g., LSTM), could be explored to explicitly capture temporal dependencies between consecutive days. The similarity-based method could be extended by giving different importance to the variables. Finally, an important extension would be the integration of these methods into real-world applications.

## 6. Ethical Considerations

The data is reused under its original conditions: participants in the original study signed an informed consent form, and the study was approved by the Institutional Review Board at Dartmouth College [4]. No new data has been collected for the present project, no participant has been contacted, and no intervention has been carried out. The data used in this project is sensitive, since variables derived from the PHQ-9, the Perceived Stress Scale, the loneliness scale and EMA mood reports relate to mental health. The GPS, WiFi location and conversation sensing modules were excluded from the analysis, and the free-text EMA responses are used only as participation indicators. The methods are not meant to be used as diagnostic or prescriptive tools, and they should not replace professional mental health support.

## 7. Conclusion

This project explored daily emotional well-being through the analysis of personalized data, with a particular focus on understanding how happiness varies across individuals in real-life conditions. The results suggest that well-being cannot be fully captured through general patterns, as individuals differ not only in their average levels of happiness, but also in their level of variability. The survey-based analysis adds a broader perspective by identifying distinct user profiles, highlighting different patterns of behaviour and well-being. At the same time, the results indicate that general predictive approaches based on behavioural and contextual data have limited ability to identify better-than-usual days. In contrast, approaches that focus on each individual's own data, such as the similarity-based method, offer a more interpretable and meaningful way to identify patterns associated with higher happiness levels.

## Acknowledgements

The author would like to thank Aleksandar Matic (Koa Health), Pablo Rodríguez (Google Research) and Alexandre Barco Martelo (University of Deusto) for their supervision and guidance throughout this work.